\documentclass[12pt,dvips]{article}
\usepackage{epsfig}

\begin{document}

\title{\bf Aspects of \\nonrelativistic quantum gravity}

\author {Johan Hansson\footnote{c.johan.hansson@ltu.se} \\
 Department of Physics, Lule{\aa} University of Technology
 \\ SE-971 87 Lule\aa, Sweden}

\date{}

\maketitle

\begin{abstract}
A nonrelativistic approach to quantum gravity is studied. At least
for weak gravitational fields it should be a valid approximation.
Such an approach can be used to point out problems and prospects
inherent in a more exact theory of quantum gravity, yet to be
discovered. Nonrelativistic quantum gravity, e.g., shows promise
for prohibiting black holes altogether (which would eliminate
singularities and also solve the black hole information paradox),
gives gravitational radiation even in the spherically symmetric
case, and supports non-locality (quantum entanglement). Its
predictions should also be testable at length scales well above
the ``Planck scale", by high-precision experiments feasible with
existing technology.
\\
\\
%PACS numbers: 03.65.-w, 03.65.Bz
\end{abstract}

%%\pacs{ PACS numbers: 03.65.-w, 03.65.Bz 11.15.-q }
%%\newpage

The greatest fundamental challenge facing physics has for many
years been to reconcile gravity with quantum physics. There have
been numerous attempts to do so, but so far there is no
established and experimentally/observationally tested theory of
``quantum gravity", the two main contenders presently being string
theory \cite{strings} and loop quantum gravity \cite{loop}, with
``outsiders" like twistor theory \cite{twistors}, non-commutative
geometry \cite{connes}, etc.

The motivations for studying nonrelativistic quantum gravity,
apart from the simple and well-defined mathematics, are:

1) Quantum theory is supposed to be universal, i.e., it should be
valid on all length scales and for all objects, as there in
principle exists no size/charge/mass-limit to its applicability.
In atomic physics the practical restriction comes about because
there is a limit to arbitrarily large atomic nuclei as, i) the
Coulomb force between protons is repulsive, eventually
overpowering the strong nuclear force trying to hold the nucleus
together, ii) the additional weak force makes neutron-rich nuclei
decay before they grow too large. Also, the electric charge comes
in both positive and negative, and as a result a big lump of
matter is almost always electrically neutral\footnote{The same
also applies for e.g. the strong force, as the three different
color charges (``red", ``green", ``blue") always combine to
produce color-neutral hadrons and bulk matter.}. Neither of these
limitations are present in ``pure" quantum gravity.

2) For weak gravitational fields the nonrelativistic theory should
be sufficient. The weak-field newtonian limit is even used for
determining the constant $\kappa$ in Einstein's field equations of
general relativity $G_{\mu \nu} = \kappa T_{\mu \nu}$. The
nonrelativistic limit is also almost always sufficient for
practical purposes in non-quantum gravity, except for a handful of
extreme cases (notably black holes and the very early universe),
although high-precision experiments in, e.g., the solar system can
and do show deviations from the nonrelativistic theory, always in
favor of general relativity \cite{Will}.

3) Even for strong gravitational fields the newtonian picture
gives the same prediction as general relativity for the
Schwarzschild radius of a spherically symmetric, non-rotating
black hole, and correct order of magnitude results for neutron
stars and cosmology. This could make it possible to deduce at
least qualitative results about strongly coupled quantum gravity,
as the nonrelativistic viewpoint should give reliable first order
quantum gravitational results.

On the other hand would any ``absurd" results obtained from
nonrelativistic quantum gravity, deviating from observations,
implicate either that:

A) General relativity cannot be quantized\footnote{This is an
automatic consequence of ``emergent" gravity, e.g. Sakharov's
theory \cite{Sakharov}, where gravity is a non-fundamental
interaction and rather a macroscopic consequence of other forces
and fields.}. An unsuccessful special case (the weak field limit)
would disprove the general case, whereas the opposite is not true.

or

B) Quantum mechanics fails at ``macroscopic" distances and for
macroscopic objects. This would mean that we in gravity have a
unique opportunity to understand the ``measurement problem" in
quantum mechanics, as proposed by e.g. K\'{a}rolyh\'{a}zy
\cite{Karolyhazy} and Penrose \cite{Penrose}. In that case we can
use gravity to probe the transition between quantum $\rightarrow$
classical behavior in detail, i.e. get experimental facts on
where, how and when the inherently undecided quantum world of
superpositions turns into the familiar objective classical
everyday world around us. Fundamental quantum gravity and the
quantum mechanical measurement problem may well be intertwined and
might need to be resolved simultaneously in a successful approach.

In nonrelativistic quantum gravity, at least as long as the system
can be approximately treated as a 2-body problem, it is possible
to use the mathematical identity between the electrostatic Coulomb
force in the hydrogen atom, and Newton's static gravitational
force under the substitution $Z e^2 / 4 \pi \epsilon_0 \rightarrow
GmM$. For weak electromagnetic fields, as in the hydrogen atom,
the electrodynamic corrections to the static Coulomb field are
very small, making the approximation excellent. The same applies
to gravity, dynamical effects from general relativity are
negligible for weak gravitational fields. A gravitationally bound
2-body system should then exhibit the same type of ``spectrum" as
a hydrogen atom, but emitted in (unobservable) graviton form
instead of photons.

For a free-falling 2-body system, e.g. in a satellite experiment
enclosed in a spherical vessel, it should in principle be possible
to measure the excitation energies for a suitable system. An
analogous result has seemingly already been accomplished for
neutrons in the gravitational field of the earth
\cite{Nezvishevsky}, although there are some quantum gravity
ambiguities as noted below.

The gravitational ``Bohr-radius", $b_0$, the innermost radius of
circular orbits in the old semi-classical Bohr-model and also the
distance $r$ for which the probability density of the
Schr\"{o}dinger equation ground-state peaks, is
\begin{equation}
b_0 = \frac{h^2}{4 \pi^2 G m^2 M} = \frac{\hbar^2}{Gm^2M},
\end{equation}
and the quantum-gravitational energy levels
\begin{equation}
E_n (grav) = - \frac{2 \pi^2 G^2 m^3 M^2}{h^2}\frac{1}{n^2} = -
\frac{G^2 m^3 M^2}{2 \hbar^2}\frac{1}{n^2} = - E_g \frac{1}{n^2},
\end{equation}
where $E_g = G^2 m^3 M^2/ 2 \hbar^2$ is the energy required to
totally free the mass $m$ from $M$ in analogy to the Hydrogen
case, whereas the expectation value for the separation is
\begin{equation}
\langle r \rangle_{grav}  \simeq n^2 b_0 = \frac{n^2 \hbar^2}{G
m^2 M}.
\end{equation}
All analytical solutions to the Schr\"{o}dinger equation, the
hydrogen wave-functions, carry over to the gravitational case with
the simple substitution $a_0 \rightarrow b_0$.
\begin{equation}
\psi_{nlm}  = R(r) \Theta(\theta) \Phi(\phi) = N_{nlm} R_{nl}
Y_{lm},
\end{equation}
where $N_{nlm}$ is the normalization constant, $R_{nl}$ the radial
wavefunction, and $Y_{lm}$, the spherical harmonics, contain the
angular part of the wavefunction.

We notice (e.g. through $b_0$) that, e.g., the planets in the
solar system must be in very highly excited quantum gravitational
states. In that sense they are analogous to electrons in ``Rydberg
atoms" in atomic physics \cite{Rydberg}.

For excited states with $l \neq 0$, and very large $n$ and $l$,
the expectation value of the distance is
\begin{equation}
\langle r \rangle \simeq \frac{1}{2}(3 n^2 - l^2) b_0,
\end{equation}
however as that is for an ensemble (average over many
measurements), for a single state it is in principle more
appropriate to use the most probable radial distance (``radius" of
orbital)
\begin{equation}
\tilde{r} = n^2 b_0,
\end{equation}
as a measure for the expected separation. However, for $n$ large
and $l=l_{max} = n-1$ the two coincide so that $\langle r \rangle
= \tilde{r}$

It is easy to show that for, e.g., Kepler's law to apply, $l$ must
be very close to $n$:

The period of revolution can be written

\begin{equation}
T = \frac{2 \pi m \tilde{r}^2}{L} = \frac{2 \pi m \tilde{r}^2}{l
\hbar},
\end{equation}
and assuming maximality for the angular momentum, $l \simeq n$,
gives

\begin{equation}
T  \simeq \frac{2 \pi m \tilde{r}^2}{n \hbar}.
\end{equation}
Solving the most probable distance, Eq. (6), for $n$ gives
\begin{equation}
n  = \frac{m \sqrt{GM\tilde{r}}}{\hbar},
\end{equation}
so that
\begin{equation}
T  \simeq \frac{2 \pi \tilde{r}^{3/2}}{\sqrt{GM}},
\end{equation}
which is Kepler's law. So, the conclusion is that all the planets
in the solar system are in maximally allowed angular momentum
states quantum mechanically. Even though the maximality of $L$ and
$L_z$ are automatic in the classical description, it is far from
obvious why the same should result from the more fundamental
quantum treatment, as noted below.

For states with $l = l_{max} = n-1$ and $m = \pm l$: i) There is
only one peak, at $r = \tilde{r}$, for the radial probability
density, and the ``spread" (variance) in the $r$-direction is
given by\footnote{The hydrogen wavefunctions for the gravitational
case give $\langle r^2 \rangle = [5n^2 + 1 -3l(l+1)] n^2 b_0^2 /2$
and $\langle r \rangle = [3n^2 - l(l+1)]b_0 /2$ .} $\Delta r =
\sqrt{\langle r^2 \rangle - \langle r \rangle^2} \simeq n^{3/2}
b_0 /2$, ii) The angular $\theta$-part of the wavefunction for
maximal $m$-quantum number $|m| = l$, is $\propto \sin^{l}\theta$.
The probability density thus goes as $sin^{2l} \theta$ in the
$\theta$-direction, meaning that only $\theta = \pi /2$ is
nonvanishing for large $l$. The azimuthal ($\phi$) part of the
angular wavefunction $Y_{lm}$ is purely imaginary, making it drop
out of the probability density, so that \textit{all} values of
$\phi$ are equally likely. (This $\phi$-symmetry is a consequence
of conservation of angular momentum in a central potential.) The
total planetary probability density is thus ``doughnut"
(torus-like) shaped, narrowly peaking around the classical
trajectory.

From a quantum gravity standpoint, the system could be in any and
all of the degenerate states, and usually at the same time, so
typical for quantum mechanical superposition. Even for given
energy and angular momentum there is no reason for objects to be
in any particular eigenstate at all of the $2l +1$ allowed, and
certainly not exclusively $m = \pm l$. The radial probability
distribution in general has $n-l$ maxima. Thus, only for $l =
l_{max} = n-1$ has it got a unique, highly peaked maximum. The
degeneracy for a given $n$ is $n^2$. Whenever $l < l_{max}$, the
radial wavefunction is highly oscillatory in $r$ as it has $n-l$
nodes. The same goes for the angular distribution as there in
general are $l-m$ nodes in the $\theta$-direction. For a general
$R_{nl} Y_{lm}$ objects could be ``all over the place", and in
simultaneous, co-existing superposed states with different quantum
numbers. Consequently, nonrelativistic quantum gravity cannot
solve the quantum mechanical measurement problem, possibly because
it lacks the non-linear terms conjectured to be needed
\cite{Penrose}.

To get the innermost allowed physical orbit for any
``test-particle", $m$, we must impose the physical restriction
that the binding energy cannot exceed the test particle
mass-energy (as the energy of the total system otherwise could not
be conserved), thus
\begin{equation}
E_g (max)  = mc^2.
\end{equation}
As $E_g$ can be written
\begin{equation}
E_g = \frac{GmM}{2 b_0},
\end{equation}
we get
\begin{equation}
b_0 (min) = \frac{GM}{2 c^2} = \frac{R_S}{4},
\end{equation}
where $R_S = 2GM/c^2$ is the Schwarzschild radius. It is amusing
to see how close $b_0 (min)$ is to $R_S$ and one cannot help
speculate that a more complete theory of quantum gravity could
ensure that $r
> R_S$ always, and thus forbid black holes altogether\footnote{For the
hydrogen atom the corresponding value is $a_0 (min) \simeq 1.4
\times 10^{-15}$ m, or one-half the ``classical electron radius",
whereas $R_S \simeq 10^{-53}$ m, so that $a_0 (min) \gg R_S$. But
we implicitly already knew that. The Coulomb force does not turn
atoms into black holes.}. In addition, all non-gravitational
radiation mechanisms during collapse have here been neglected. The
object $M$ must be put together somehow, but if $r_{min}
> R_S$ it can never accrete enough matter to become a black hole,
as the infalling mass (energy) instead will be radiated away in
its totality (in gravitons), making a black hole state impossible
\cite{Hansson}. This would, in an unexpected way, resolve the
black hole information loss paradox. Even though $r=R_S$
represents no real singularity, as it can be removed by a
coordinate transformation, anything moving inside $r < R_S$ will,
according to classical general relativity, in a (short) finite
proper time reach the true singularity at $r=0$. If quantum
gravity could ensure that $r > R_S$ always, gravity would be
singularity free.

Let us briefly look at radiative transitions. From the dipole
approximation in atomic physics an elementary quantum (photon)
transition requires $\Delta l = \pm 1$. A quadrupole (graviton)
approximation in quantum gravity instead requires $\Delta l = \pm
2$. So, a typical elementary energy transfer in a highly excited,
gravitationally bound 2-body quantum gravitational system is
\begin{equation}
\Delta E  = - E_g (n^{-2} - (n-2)^{-2}) \simeq \frac{4 E_g}{n^3}.
\end{equation}

We also see that the gravitational force is not really
conservative, even in the static newtonian approximation. The
changes in kinetic and potential energies do not exactly balance,
$\Delta K \neq \Delta U$, the difference being carried away by
gravitons in steps of $\Delta l = 2$. Also, in the quantum gravity
case there is gravitational radiation even in the spherically
symmetric case, which is forbidden in classical general
relativity.

We are now equipped to return to the experiment with neutrons in
the gravitational field of the earth \cite{Nezvishevsky}, claiming
to have seen, for the first time, quantum gravitational states in
the potential well formed by the approximately linear
gravitational potential near the earth surface and a horizontal
neutron mirror. An adjustable vertical gap between the mirror and
a parallel neutron absorber above was found to be non-transparent
for traversing neutrons for separations less than $\sim 15 \mu$m
(essentially due to the fact that the neutron ground state
wavefunction then overlaps the absorber). As the neutron in such a
well, from solving the Schr\"{o}dinger equation, has a ground
state wavefunction peaking at $\sim 10 \mu$m, with a corresponding
energy of $\simeq 1.4 \times 10^{-12}$ eV, the experimental result
is interpreted to implicitly having verified, for the first time,
a gravitational quantum state.

If we instead analyze the experiment in the framework of the
present article, the same experimental setup gives $b_0 \simeq 9.5
\times 10^{-30}$ m, $E_g \simeq 2.2 \times 10^{35}$ eV. Close to
the earth's surface, $\tilde{r} \simeq R_{\oplus} \simeq 6.4
\times 10^6$ m, the radius of the earth, giving $n \simeq 8.2
\times 10^{17}$, resulting in a typical energy for an elementary
quantum gravity transition $\Delta E \simeq 4 E_g /n^3 \simeq 1.6
\times 10^{-18}$ eV. For a cavity of $\Delta \tilde{r} = 15 \mu$m,
and $n \gg \Delta n \gg 1$, one gets $\Delta E = E_g b_0 \Delta
\tilde{r} /\tilde{r}^2 \simeq 0.7 \times 10^{-12}$ eV = 0.7 peV,
to be compared to the value 1.4 peV as quoted in
\cite{Nezvishevsky}. Even though the present treatment gives a
similar value for the required energy, it need \textit{not} be the
result of a \textit{single} quantum gravity state as calculated in
\cite{Nezvishevsky}, but rather $\leq 10^6$ gravitons can be
emitted/absorbed. From the treatment in this article it is thus
not self-evident to see why the experimental apparatus
\cite{Nezvishevsky} should be non-transparent to neutrons for
vertical separations $\Delta h < 15 \mu$m.

% LL-, LS-coupling??
%
%TRANSITIONS? (LIFETIMES)? i) Dipole (Rydberg)?, ii) Quadrupole?
%(graviton)
%
%COLLELA-OVERHAUSER-WERNER ("COW") EXPERIMENT???
%
%POUND-REBKA EXPERIMENT???
%
%TITIUS-BODE'S LAW???

Thus, the difference regarding quantized energy levels for an
experiment with neutrons ``falling"\footnote{In fact, a bound
quantum gravitational object does not fall at all as it is
described by a stationary wavefunction, or a superposition of
such.} under the influence of earth's gravity \textit{with} mirror
(as in \cite{Nezvishevsky}) or \textit{without} (above) shows that
nonrelativistic quantum gravity is dependent on global boundary
conditions, where the boundary in principle can lie arbitrarily
far away. This comes as no surprise, as the Schr\"{o}dinger
equation models the gravitational interaction as instantaneous,
contrasted with the case in general relativity where the behavior
in free-fall only depends on the local properties of mass-energy
and the resulting spacetime curvature (out of which the mirror is
not part due to its inherently non-gravitational interaction with
the neutron) and causal connection as the gravitational
interaction propagates with the speed of light. However, as
several experiments on entangled quantum states, starting with
Clauser/Freedman \cite{Clauser} and Aspect et al. \cite{Aspect},
seem to be compatible with a non-local connection between quantum
objects \cite{Bell}, this property of the Schr\"{o}dinger equation
need not be a serious drawback for a theory of quantum gravity.

%\begin{figure}[h]
%\begin{center} \psfig{file=fig1.eps}
%\leavevmode
%\epsffile{logistic.eps}
%\end{center}

% \caption{}
% \end{figure}

\end{document}